\RequirePackage{fixltx2e}
\documentclass[reprint, amssymb, amsmath, amsfonts, aip, cha, floatfix, showpacs, twoside]{revtex4-1}

\usepackage{xspace, enumerate, graphicx, listings, mathtools}

\newcommand{\abs}[1]{\mathopen{}\mathclose\bgroup\left\lvert#1\aftergroup\egroup\right\rvert}
\newcommand{\norm}[1]{\mathopen{}\mathclose\bgroup\left\|#1\aftergroup\egroup\right\|}
\newcommand{\kl}[1]{\mathopen{}\mathclose\bgroup\left(#1\aftergroup\egroup\right)}
\newcommand{\klg}[1]{\mathopen{}\mathclose\bgroup\left\{#1\aftergroup\egroup\right\}}
\newcommand{\kle}[1]{\mathopen{}\mathclose\bgroup\left[#1\aftergroup\egroup\right]}
\newcommand{\kls}[1]{\mathopen{}\mathclose\bgroup\left\langle#1\aftergroup\egroup\right\rangle}
\newcommand{\floor}[1]{\mathopen{}\mathclose\bgroup\left\lfloor#1\aftergroup\egroup\right\rfloor}
\newcommand{\ceil}[1]{\mathopen{}\mathclose\bgroup\left\lceil#1\aftergroup\egroup\right\rceil}
\newcommand{\leftopen}[1]{\mathopen{}\mathclose\bgroup\left(#1\aftergroup\egroup\right]}
\newcommand{\rightopen}[1]{\mathopen{}\mathclose\bgroup\left[#1\aftergroup\egroup\right)}

\newcommand{\defi}{\mathrel{\mathop:}=}
\newcommand{\ifed}{=\mathrel{\mathop:}}

\newcommand*\diff{\mathop{}\!\mathrm{d}}

\newcommand{\Epi}{\affiliation{Department of Epileptology, University of Bonn, Sigmund-Freud-Stra{\ss}e~25, 53105~Bonn, Germany}}
\newcommand{\HISKP}{\affiliation{Helmholtz Institute for Radiation and Nuclear Physics, University of Bonn, Nussallee~14--16, 53115~Bonn, Germany}}
\newcommand{\IZKS}{\affiliation {Interdisciplinary Center for Complex Systems, University of Bonn, Br\"uhler Stra\ss{}e~7, 53175~Bonn, Germany}}

\begin{document}

\title{Efficiently and easily integrating differential equations with JiTCODE, JiTCDDE, and JiTCSDE}

\author{Gerrit Ansmann}
\Epi \HISKP \IZKS

\begin{abstract}
We present a family of Python modules for the numerical integration of ordinary, delay, or stochastic differential equations.
The key features are that the user enters the derivative symbolically and it is just-in-time-compiled, allowing the user to efficiently integrate differential equations from a higher-level interpreted language.
The presented modules are particularly suited for large systems of differential equations such as used to describe dynamics on complex networks.
Through the selected method of input, the presented modules also allow to almost completely automatize the process of estimating regular as well as transversal Lyapunov exponents for ordinary and delay differential equations.
We conceptually discuss the modules' design, analyze their performance, and demonstrate their capabilities by application to timely problems.

\end{abstract}

\maketitle

\begin{quotation}
	Solving differential equations is an integral part of many simulation studies in various scientific fields ranging from physics over neuroscience to economics as well as of the theoretical investigations of dynamical systems.
	As this task can often only be solved numerically, there is a wide need for software dedicated to it.
	Two relevant but often conflicting criteria for the design of such a software are usability and efficiency.
	The latter is of particular interest for the integration of large systems of differential equations as they arise in the simulation of dynamics on complex networks.
	We here present a family of software modules that follow a new design approach, which requires only little trade-off between efficiency and usability, and allows to automatize many tedious and error-prone steps on behalf of the user.
\end{quotation}

\section{Introduction}

While the theory of dynamical systems allows to analytically derive some properties of the solutions to differential equations~\cite{strogatz1994, alligood1996}, the solutions themselves can often only be approximated numerically.
Algorithms for solving ordinary differential equations have been investigated intensively and a plethora of implementations of performant algorithms are available, in particular for ordinary differential equations (ODEs)~\cite{hairer1987, deuflhard2002}, but also for the more difficult problems of delay differential equations (DDEs)~\cite{shampine2001, bellen2003} and stochastic differential equations (SDEs)~\cite{kloeden1999, roessler2010}.

However, performance of these algorithms can only be measured in the number of evaluations of the derivative (and the diffusion term for SDEs), i.e., the right-hand side of the differential equation.
This task depends on the individual problem and is often not generalizable.
(An important exception to this are partial differential equations, which we do not consider here as an entire discipline devoted to these has already yielded specialized algorithms and tools.)
In most existing performant software designs and workflows, the user specifies the derivative in a lower-level language, and then they either have to compile it themselves or it is compiled on their behalf by the software (see e.g., Ref.~\onlinecite{clewley2007}).
In particular when working with a higher-level programming language, this requires the user to switch to another language.
Moreover, it is unfeasible to explicitly specify large systems of differential equations, which in particular occur during the simulation of dynamics on complex networks~\cite{Boccaletti2006, Arenas2008}.
For the latter, it is possible to evaluate the network structure at execution time~\cite{Rothkegel2012}, but this comes with an inevitable loss of performance and implementing non-predefined node dynamics or couplings may prove difficult.

We here present JiTCODE, JiTCDDE, and JiTCSDE, a family of Python~\cite{oliphant2007} modules for integrating ordinary, delay, and stochastic differential equations, respectively.
In contrast to existing modules, the user specifies the derivative symbolically within Python, and thus can employ Python's control structures and other features when doing so, which for example makes it straightforward to specify the differential equations for a complex network.
These symbolic differential derivatives are then automatically translated to C~code, compiled, and loaded by Python again without requiring the user to provide any input or use another language.
The presented modules thus allow the user to enjoy both, the efficiency of a lower-level compiled language and the ease of use of a higher-level interpreted language.
Moreover the symbolic input allows to automatize further processing of the derivative such as needed for estimating Lyapunov exponents and for certain implicit integrators.

This paper is mostly split into conceptual and example sections, which have little dependence on each other, such that it is possible to skip sections of one kind.
Basic Python features will not be explained in the example sections and we trust Python neophytes to look them up or infer their usage from the examples.
On the other hand, experienced Python users will notice that we mostly avoid using more complex Python structures.
Finally note that for didactic reasons but against good practice, we use ``last-minute imports''.
Source files for all examples are available as ancillary files.
The presented modules are available on \url{https://github.com/neurophysik} and from the Python Package Index and can usually be installed with something along the lines of \texttt{pip3 install jitcode}.
This paper is based on version~1.2 of each of the presented modules, except for the benchmark shown in Sec.~\ref{benchmark}, which was performed using version~1.0.

\section{General design and rationale}

We consider the following three types of differential equations:
\begin{itemize}
	\item Ordinary differential equations (ODEs):
		\begin{equation}\label{eq:ode}
			\dot{y}(t) = f\kl{t, y(t)},
		\end{equation}
		where \(y: \mathbb{R} \rightarrow \mathbb{R}^n\) and \(f: \mathbb{R} \times \mathbb{R}^n \rightarrow \mathbb{R}^n\).
	
	\item Delay differential equations (DDEs):
		\begin{equation}\label{eq:dde}
			\dot{y}(t) = f \big ( t, y(t), y(t-\tau_1), y(t-\tau_2), \ldots \big ),
		\end{equation}
		where \(y: \mathbb{R} \rightarrow \mathbb{R}^n\) and \(f: \mathbb{R} \times \mathbb{R}^n \times \mathbb{R}^n \times \ldots \rightarrow \mathbb{R}^n\).
	
	\item Stochastic differential equations (SDEs) of It\=o or Stratonovich type:
		\begin{equation}\label{eq:sde}
		\diff y(t) = f(t,y(t)) \diff t + g(t,y(t)) \odot \diff W(t),
		\end{equation}
		where \(y: \mathbb{R} \rightarrow \mathbb{R}^n\), \(f, g: \mathbb{R} \times \mathbb{R}^n \rightarrow \mathbb{R}^n\), \(W\) is an \(n\)\nobreakdash-dimensional standard Wiener process, and \(\odot\) denotes the component-wise multiplication of vectors.
\end{itemize}
With respect to performance, we focus on the case that \(n\)~is large and \(f\) has no structure that can be easily vectorized, which in particular applies to complex networks.

\newcommand{\Order}[2][n]{\ensuremath{\mathcal{O}(#2) \text{ as } #1 \rightarrow \infty}}

\label{rationale}

When numerically integrating such differential equations, there are two main computational time sinks (that are not \Order{1}):
The first is performing standard operations required by the method of integration, such as vector arithmetics or the generation of random numbers.
These usually scale with \Order{n}.
As they do not depend on the specific differential equation, can easily be vectorized or performed with existing, highly optimized routines, there is little room for improvement here.
The second and more challenging time sink is evaluating the derivative~\(f\) and possibly the Jacobian of~\(f\), which scales with \Order{n} or worse.
(For SDEs, this also includes evaluating the diffusion strength~\(g\), for which the following considerations and procedures for~\(f\) apply analogously.)
When computing dynamics on a network, the main time sink within this often is computing the coupling term;
for a high average degree, it may dominate the entire runtime.
If the network is sparse, this can be implemented with an adjacency list~\cite{Rothkegel2012}, which is equivalent to a sparse matrix--vector multiplication for linear couplings; however this inevitably causes some overhead through look-ups.

\begin{figure}
	\includegraphics[width=\linewidth]{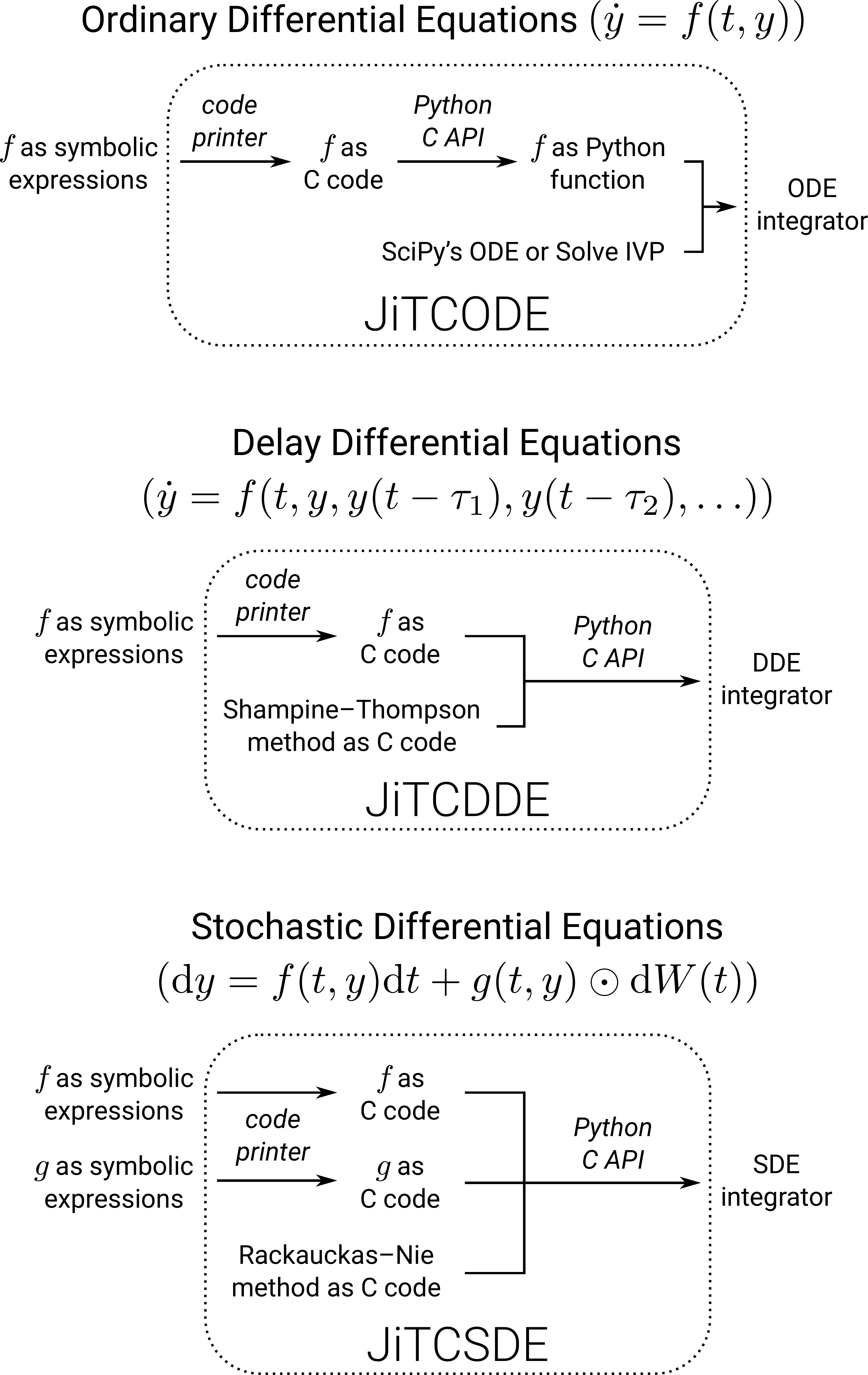}
	\caption{Basic internal structure of the three modules presented in this paper.}
	\label{fig:structure}
\end{figure}

One way to avoid this is to hard-code the entire evaluation of~\(f\).
However for large networks, directly writing the respective code is highly unfeasible, and thus some form of metaprogramming must be employed.
The modules presented in this paper provide a framework in Python that handles this metaprogramming for the user.
Their general internal workflow is the following (see also Fig.~\ref{fig:structure}):
\begin{enumerate}
	\item The user provides \(f\) in symbolic form, more specifically as a list or generator function (see App.~\ref{generators}) of SymEngine~\cite{meurer2017} expressions.
		SymEngine --~more accurately: its Python bindings~-- is a fast symbolic module for Python.
		It is aspiring to replace the pure Python core of the better known SymPy~\cite{meurer2017}, and thus the two are almost identical in handling and compatible with each other.
	\item Using SymEngine's code-printing routines, these expressions are converted to C~code.
		\pagebreak
	\item The C~code is embedded in a template that makes it a C~extension of Python.
	\item This C~extension is then compiled and loaded by Python, resulting in an efficient evaluation of~\(f\) accessible from Python.
	\item This is combined with an integrator and provided to the user.
		For ODEs an existing integration module for Python is employed; for DDEs and SDEs, the integrator is provided as a C~implementation which is compiled together with~\(f\).
\end{enumerate}
By default, these steps are fully automatized and do not require any input from the user.
If desired, they can be tweaked, e.g., by choosing different compilation parameters.

The symbolic input has some further advantages:
\begin{itemize}
	\item It allows for an intuitive user-interface:
		The differential equations can be entered almost as they are written on paper, using Python's standard arithmetic operations and control structures.
	\item \(f\) can be further processed automatically, if needed.
		For example, it is possible to employ symbolic differentiation to convert Stratonovich to It\=o SDEs~(see Sec.~\ref{jitcsde}) or to obtain the Jacobian of~\(f\) as needed for calculating Lyapunov exponents (see Sec.~\ref{lyapunov}) or as needed by some integrators (see Sec.~\ref{jitcode}).
	\item Some trivial optimizations can be performed automatically, e.g., summands that are zero are automatically discarded.
		Moreover, symbolic simplification as well as common-subexpression elimination can be employed.
	\item The user can re-use the input for their own symbolic analysis of the differential equation, e.g., to determine fixed points.
\end{itemize}

Apart from Python~3, a C~compiler, and SymEngine, the presented modules only depend on Setuptools, NumPy, Jinja~2, and SciPy (only JiTCODE).
The latter are highly popular modules that are commonly available and can be expected to remain maintained and be backwards-compatible for the foreseeable future.
Given these requirements, we confirmed that the compilation backend of the presented modules works out of the box on several Linux distributions, FreeBSD, MacOS, as well as on Windows (in an Anaconda environment using the MSVC compiler).
Should compilation not be possible, the presented modules can also completely operate within Python (except for employed modules like NumPy and SymEngine for which binaries are readily available) as a fallback, which, while considerably slower, may suffice for an analysis of small systems of differential equations.

\section{The three modules}

\subsection{JiTCODE -- Just-in-Time Compilation for Ordinary Differential Equations}\label{jitcode}

JiTCODE is designed for ordinary differential equations as specified in Eq.~\ref{eq:ode}.
As the evaluation of the derivative~\(f\) and the integration procedure are not intricately intertwined, it is possible to rely on existing software for the latter, namely the SciPy~\cite{libscipy,oliphant2007} modules \texttt{scipy.integrate.ode} and \texttt{scipy.integrate.solve\char`_ivp} (see also Fig.~\ref{fig:structure}, top).
As of now, these modules provide seven integration schemes:
Dormand's and Prince's fifth-order method~\cite{dormand1980,hairer1987},
the DoP853 method~\cite{hairer1987},
Bogacki’s and Shampine’s third-order method~\cite{bogacki1989},
VODE~\cite{brown1989},
LSODA~\cite{hindmarsh1983},
a fifth-order Radau IIA method~\cite{hairer1996},
and an implicit multi-step variable-order method based on a backward differentiation formula~\cite{shampine1997}.
The latter four can make use of the Jacobian, which JiTCODE will automatically determine using symbolic differentiation, render as C~code and compile (like the derivative~\(f\)) if such a solver is selected.
The SciPy modules are interfaced such that future improvements of the employed integrators and newly added integrators will likely be available to JiTCODE automatically or with little adaption.

\subsubsection*{Example: R\"ossler oscillator}\label{roessler}

We want to integrate a R\"ossler oscillator~\cite{roessler1976}, governed by the following differential equations:
\begin{equation}
	\begin{alignedat}{3}\label{eq:roessler}
		\dot{y}_0 &=& \,-y_1 &-   y_2, \\
		\dot{y}_1 &=&    y_0 &+ a y_1, \\
		\dot{y}_2 &=&      b &+ y_2 (y_0 - c),
	\end{alignedat}
\end{equation}
with the control parameters
\(a = 0.2\),
\(b = 0.2\), and
\(c = 5.7\).

\begin{figure*}
	\includegraphics{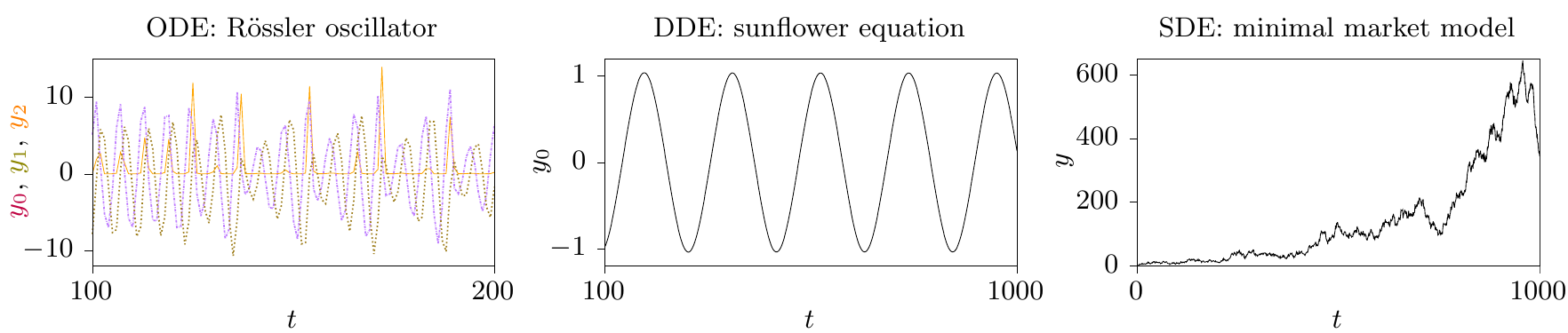}
	\caption{Outputs (time series) of the first three example programs.}
	\label{fig:timeseries}
\end{figure*}

We now set up this system with JiTCODE and start by defining our control parameters as regular numbers:
\begin{lstlisting}[firstline=1,lastline=4]
a = 0.2
b = 0.2
c = 5.7
\end{lstlisting}
We then import the symbolic function~\texttt{y}, which represents the dynamical variables, and use it to implement the right-hand side of Eq.~\ref{eq:roessler} as a list \texttt{roessler\char`_f}:
\begin{lstlisting}[firstline=6,lastline=11]
from jitcode import y
roessler_f = [
		-y(1) - y(2),
		 y(0) + a*y(1),
		b + y(2)*(y(0)-c)
	]
\end{lstlisting}
\pagebreak
Note that, as \texttt{y} is a SymEngine symbolic function, the elements of the list will automatically be SymEngine expressions.
We can use such a list to instantiate the \texttt{jitcode} class, obtaining an integrator object~\texttt{I}:
\begin{lstlisting}[firstline=19,lastline=20]
from jitcode import jitcode
I = jitcode(roessler_f)
\end{lstlisting}

The object~\texttt{I} can now be used to set up and solve initial-value problems (it behaves like an instance of \texttt{scipy.integrate.ode}).
First, we choose an integrator, namely the adaptive fifth-order method by Dormand and Prince (\texttt{dopri5}):
\begin{lstlisting}[firstline=22,lastline=22]
I.set_integrator("dopri5")
\end{lstlisting}
Then we choose the initial condition and time; in this case we start at \(t=0\) with
\(y_0(0) = 0.1\),
\(y_1(0) = 0.2\), and
\(y_2(0) = 0.3\):
\begin{lstlisting}[firstline=24,lastline=25]
initial = [0.1,0.2,0.3]
I.set_initial_value(initial,time=0.0)
\end{lstlisting}
Note that this will trigger the automatic compilation of the derivative.
Finally, we want to integrate up to \(t=200\) and \texttt{print} the state of the system for \(t \in \{100,101,\ldots,200\}\), thereby ignoring the first \(100\) time units to let transients die out.
For this purpose, we use that \texttt{integrate} returns the state of the system after integration:
\begin{lstlisting}[firstline=27,lastline=29]
times = range(100,201)
for time in times:
	print(I.integrate(time))
\end{lstlisting}
While we here \texttt{print} the states to keep the example simple, they can of course be further processed within Python, e.g., for analysis and plotting.
The output is plotted in Fig.~\ref{fig:timeseries}, left.
\pagebreak
\subsection{JiTCDDE -- Just-in-Time Compilation for Delay Differential Equations}\label{jitcdde}
JiTCDDE is designed for delay differential equations (DDEs) as described in Eq.~\ref{eq:dde}:
In contrast to JiTCODE, no existing module for DDE integration is used, since DDEs require the evaluation of~\(f\) and the actual integration routine to be more intertwined.
JiTCDDE employs the method proposed by Shampine and Thompson~\cite{shampine2001}, which is based on Bogacki's and Shampine's third-order method~\cite{bogacki1989} and also implemented in the Matlab module~\texttt{dde23}.
We chose this method since to our knowledge it is of the highest order where the interpolation error (when obtaining the past) is of the same order as the integration error of an explicit method~\cite{shampine2001}.
JiTCDDE contains a C~implementation of this method, which is compiled together with the derivative~\(f\) into a C~extension of Python, which in turn is provided to the user (see central panel of Fig.~\ref{fig:structure}).

The method features an adaptive step size and accesses past states through a piecewise cubic Hermite interpolation (or extrapolation in some special cases) of the states and derivatives at previous time steps \textit{(anchors).}
To store and access these anchors, we make use of a doubly linked list with a cursor (see App.~\ref{past_access}).
To avoid having to treat the initial past separately, it is stored as such a list of anchors as well.
While the user can provide the initial past in this format, JiTCDDE also provides convenience functions that determine anchors heuristically from a function supplied by the user or just set the initial past to a constant value.
Storing the initial past this way also has the advantage that the only discontinuity that requires handling is that of the derivative at the start time of the integration.
To this purpose, two methods are provided:
One uses discontinuity tracking as proposed by Shampine and Thompson~\cite{shampine2001}; the other integrates with fixed time steps for a while, ignoring the error estimate.

JiTCDDE has no technical restrictions against state- or time-dependent delays.
On the one hand, such delays may warp the error estimate for purposes of the adaptive integration, as it does not take into account the inaccuracy caused by a changing delay.
On the other hand, this should not be an issue if the delays change sufficiently slowly in comparison to the step size.

\pagebreak

\subsubsection*{Example: sunflower equation}\label{sunflower}

We want to integrate the sunflower equation~\cite{somolinos1978}:
\begin{equation}
	\begin{alignedat}{1}\label{eq:sunflower}
		\dot{y}_0(t) &= y_1(t),\\
		\dot{y}_1(t) &= 
			- \frac{a}{\tau}y_1(t)
			- \frac{b}{\tau} \sin\kl{y_0(t-\tau)},
	\end{alignedat}
\end{equation}
with the control parameters
\(\tau = 40\),
\(a = 4.8\), and
\(b = 0.186\).
Again we start by defining our control parameters:
\begin{lstlisting}[firstline=1,lastline=3]
a = 4.8
b = 0.186
tau = 40
\end{lstlisting}
In contrast to the previous example, we also have to import a symbol for the time as well as the symbolic sine function (note that we cannot use \texttt{math.sin} or \texttt{numpy.sin} here since they cannot handle symbolic input):
\begin{lstlisting}[firstline=5,lastline=6]
from jitcdde import y, t
from symengine import sin
\end{lstlisting}
We can then implement the right-hand side of Eq.~\ref{eq:sunflower} as a list and instantiate the \texttt{jitcdde} class:
\begin{lstlisting}[firstline=8,lastline=13]
sunflower_f = [
		y(1),
		-a/tau*y(1) - b/tau*sin(y(0,t-tau))
	]
from jitcdde import jitcdde
I = jitcdde(sunflower_f)
\end{lstlisting}
We want the initial past to be
\(y_0(\tilde{t}) = 1.0\) and
\(y_1(\tilde{t}) = 0.0\)
for \(\tilde{t}<0\).
We can thus use the most simple of several ways to initiate the past:
\begin{lstlisting}[firstline=15,lastline=15]
I.constant_past( [1.0,0.0], time=0.0 )
\end{lstlisting}
Before we start with the actual integration, we have to address initial discontinuities, which we do by discontinuity tracking~\cite{shampine2001}.
The following command automatically performs this, i.e., it performs steps to match points where \(f\)~is not sufficiently smooth:
\begin{lstlisting}[firstline=16,lastline=16]
I.step_on_discontinuities()
\end{lstlisting}
Alternatively we could have called \texttt{I.integrate\char`_blindly(tau)} or with a higher argument to ignore the estimate of the integration error until the initial discontinuities have been sufficiently smoothed.
Finally, we can integrate and print the output as in the previous example, just that this time we are only interested in the component~\(y_0\) (plotted in the central panel of Fig.~\ref{fig:timeseries}):
\begin{lstlisting}[firstline=18,lastline=20]
times = range(100,1001)
for time in times:
	print(I.integrate(time)[0])
\end{lstlisting}

\subsection{JiTCSDE -- Just-in-Time Compilation for Stochastic Differential Equations}\label{jitcsde}

JiTCSDE is designed for stochastic differential equations (SDEs) as described in Eq.~\ref{eq:sde}:
It employs the adaptive method for It\=o SDEs proposed by Rackauckas and Nie~\cite{rackauckas2017}, which in turn employs two embedded R\"o\ss{}ler-type stochastic Runge--Kutta methods~\cite{roessler2010}.
We chose this method since to our knowledge it is the most feasible adaptive method for SDEs.
JiTCSDE contains a C~implementation of this method, which is compiled together with the drift and diffusion functions \(f\) and~\(g\) into a C~extension of Python, which in turn is provided to the user (see the bottom of Fig.~\ref{fig:structure}).

The user can also specify the SDE in Stratonovich calculus, in which case JiTCSDE automatically applies the drift-conversion formulas~\cite{cyganowski2001} to obtain an It\=o SDE as required by the integrator.
In case of additive noise, i.e. \(\frac{\partial{g}}{\partial y}=0\), a simpler and faster variant of the method is employed.
As the input is symbolic, this can be decided completely automatically.

JiTCSDE also provides means to integrate Eq.~\ref{eq:sde} extended by jumps~\cite{merton1976} (only for It\=o SDEs).
This functionality is implemented purely in Python, which makes it very flexible, but also potentially slow.
However, the latter should not have a relevant impact on the overall runtime if the jumps are rare in comparison to the integration step and if the SDE is high-dimensional.

\subsubsection*{Example: minimal market model}
We want to integrate the stylized minimal market model (MMM)~\cite{platen2010} described by the following It\=o SDE:
\begin{equation}\label{eq:MMM}
	\diff y = \alpha \exp\kl{\eta t} \diff t
			  + \sqrt{\alpha \exp\kl{\eta t} y} \diff W
\end{equation}
with the control parameters \(\alpha=0.2\) and \(\eta=0.001\).
After setting the control parameters, we have to define the drift and diffusion function separately.
As this process is one-dimensional, we only refer to the state with the index~\texttt{0} and use lists of length~\(1\):
\begin{lstlisting}[firstline=1,lastline=7]
alpha = 0.2
eta = 0.001

from jitcsde import y, t
from symengine import exp, sqrt
MMM_f = [ alpha*exp(eta*t) ]
MMM_g = [ sqrt( alpha*exp(eta*t)*y(0) ) ]
\end{lstlisting}
We then use the drift and diffusion function to instantiate the \texttt{jitcsde} class and set the initial condition to \(y(0)=1.0\):
\begin{lstlisting}[firstline=9,lastline=11]
from jitcsde import jitcsde
I = jitcsde(MMM_f,MMM_g,ito=True)
I.set_initial_value([1.0],time=0.0)
\end{lstlisting}
Finally, as in the previous examples, we integrate and print the output (plotted in the right panel of Fig.~\ref{fig:timeseries}):
\begin{lstlisting}[firstline=16,lastline=18]
times = range(0,1001)
for time in times:
	print(I.integrate(time))
\end{lstlisting}

\section{Features and Capabilities}

\subsection{Networks}

For the low-dimensional examples shown so far, we did not fully use the metaprogramming capabilities of the presented modules --~because we did not need to:
Our examples allowed us to specify the differential equations as explicit lists of symbolic expressions.
However, for this purpose, we only used Python's native syntax, operations, and data structures.
Therefore, implementing networks is straightforward and only requires simple Python tools such as loops and sums.

One feature of Python that is particularly useful for implementing networks are generator functions.
The presented modules accept these as an alternative input format besides lists, which we make use of in the following example.
For readers who are unfamiliar with generator functions, we give a brief introduction in App.~\ref{generators}.

\subsubsection*{Example: random network of Kuramoto oscillators} \label{kuramotos}

We want to integrate \(n=100\) Kuramoto oscillators~\cite{kuramoto1975}, with the \(i\)\textsuperscript{th} oscillator being described by the following differential equation:
\begin{equation}\label{eq:kuramoto}
	\dot{y}_i = \omega_i + \frac{c}{n-1} \sum_{j=0}^{n-1} A_{ji} \sin\kl{y_j-y_i},
\end{equation}
where \(\omega_i \sim \mathcal{U} \kl{\kle{-0.5,0.5}}\) is the oscillator's eigenfrequency, \(c=3.0\) is the coupling constant, and \(A \in \klg{0,1}^{n \times n}\) is the adjacency matrix of a directed random network, in which each edge exists with a probability \(q=0.2\).

We start with parameter definitions:
\begin{lstlisting}[firstline=1,lastline=3]
n = 100
c = 3.0
q = 0.2
\end{lstlisting}
To generate the adjacency matrix \(A\), we employ the \texttt{choice} function from NumPy's \texttt{random} module to generate an \(n \times n\) array, where each element is \texttt{1} with a probability of~\(q\) and \texttt{0} with a probability of~\(1-q\):
\begin{lstlisting}[firstline=5,lastline=6]
from numpy.random import choice
A = choice( [1,0], size=(n,n), p=[q,1-q] )
\end{lstlisting}
(As \(\sin\kl{y_i-y_i}=0\), self-couplings do not contribute to the coupling sum anyway, and thus it does not matter that we also fill the diagonal with random values.)
While such a random network can be easily realized, other coupling topologies usually require more complicated user-written functions or the use of dedicated tools~\cite{csardi2006, hagberg2008} to instantiate network models.
Next we create an array of size~\(n\) containing the eigenfrequencies:
\begin{lstlisting}[firstline=8,lastline=9]
from numpy.random import uniform
omega = uniform(-0.5,0.5,n)
\end{lstlisting}
As the network nodes are statistically identical, we can sort this array to enhance the readability of a resulting plot (Fig.~\ref{fig:kuramotos}):
\begin{lstlisting}[firstline=11,lastline=11]
omega.sort()
\end{lstlisting}
With this we have everything we need to implement the right-hand side of Eq.~\ref{eq:kuramoto} for all~\(i \in \klg{0,\ldots,n-1}\) as a generator function (see App.~\ref{generators}):
\begin{lstlisting}[firstline=13,lastline=22]
from jitcode import y
from symengine import sin
def kuramotos_f():
	for i in range(n):
		coupling_sum = sum(
				sin(y(j)-y(i))
				for j in range(n)
				if A[j,i]
			)
		yield omega[i] + c/(n-1)*coupling_sum
\end{lstlisting}
Note that we could be closer to Eq.~\ref{eq:kuramoto} by writing ``\texttt{A[j,i]*sin(y(j)-y(i))}'' and removing the line ``\texttt{if A[j,i]}''.
However, this would lead to a slight slow-down of the code generation, as \texttt{sin(y(j)-y(i))} would also have to be evaluated for non-existing edges.

We can now initialize \texttt{jitcode}, and choose the integrator.
In contrast to the previous examples, we here explicitly pass the dimension~(\texttt{n}) of the differential equation to avoid that JiTCODE has to spend time to determine it on its own by iterating over the generator function.
Moreover, we set the integrator to only use an absolute error criterion for adjusting the step size, as it makes more sense given the cyclic nature of the Kuramoto oscillators.
\begin{lstlisting}[firstline=34,lastline=36]
from jitcode import jitcode
I = jitcode(kuramotos_f,n=n)
I.set_integrator("dopri5",atol=1e-6,rtol=0)
\end{lstlisting}
We want the integration to start at \(t=0\) and the initial states to be i.i.d. random phases, i.e., \(y_i(0) \sim \mathcal{U}\kl{\rightopen{0,2\pi}}\):
\begin{lstlisting}[firstline=38,lastline=40]
from numpy import pi
initial_state = uniform(0,2*pi,n)
I.set_initial_value(initial_state,time=0.0)
\end{lstlisting}

Lastly, we perform the actual integration and \texttt{print} the results after mapping them to the interval~\(\rightopen{0,2\pi}\) (plotted in Fig.~\ref{fig:kuramotos}):
\begin{lstlisting}[firstline=42,lastline=44]
times = range(0,2001)
for time in times:
	print( I.integrate(time) % (2*pi) )
\end{lstlisting}

\begin{figure}
	\includegraphics{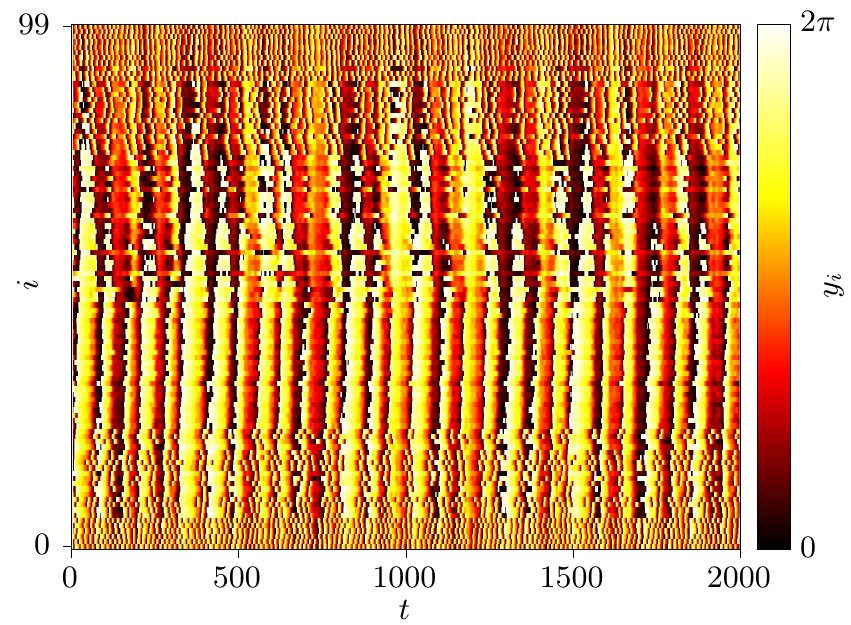}
	\caption{Output of the example program from Sec.~\ref{kuramotos}, which integrates a random network of Kuramoto oscillators.
	The oscillators are sorted by their eigenfrequencies.}
	\label{fig:kuramotos}
\end{figure}

\subsection{Lyapunov exponents}\label{lyapunov}
JiTCODE and JiTCDDE provide means to easily estimate the Lyapunov exponents for a given ODE or DDE from the evolution of tangent vectors.
For ODEs, we implement the approach by Benettin et al.~\cite{benettin1980}.
For DDEs, we mainly follow Farmer's adaption of this approach from ODEs to DDEs~\cite{farmer1982}, but extend it to handle adaptive step sizes by employing a function scalar product (see App.~\ref{DDE_lyap}).

More specifically, the following steps are automatized:
\begin{itemize}
	\item Symbolically calculating the partial derivatives required to obtain the differential equations for the tangent vectors.
	\item Implementing these differential equations alongside the differential equations for the main dynamics.
	\item Initializing the tangent vectors with randomly oriented vectors of length~1.
	\item Obtaining the norms of the tangent vectors and (ortho)normalizing them.
	\item Calculating the local Lyapunov exponents for each sampling step.
\end{itemize}
All that is left to the user is giving the tangent vectors time to align and apply the appropriate statistics, in particular averaging, to the local Lyapunov exponents.

JiTCODE also provides the tangent vectors (Lyapunov vectors) to the user.

\pagebreak

\subsubsection*{Example: Lyapunov exponents of the sunflower equation} \label{sunflower_lyap}
We want to determine the three largest Lyapunov exponents \(\lambda_1\), \(\lambda_2\), and~\(\lambda_3\) of the sunflower equation and re-use \texttt{sunflower\char`_f} as defined before (see Sec.~\ref{sunflower}).
Instead of \texttt{jitcdde}, we use \texttt{jitcdde\char`_lyap} to initialize the integrator and specify that we wish to calculate three Lyapunov exponents:
\begin{lstlisting}[firstline=22,lastline=23]
from jitcdde import jitcdde_lyap
I = jitcdde_lyap(sunflower_f,n_lyap=3)
\end{lstlisting}
This step automatically generates the differential equations for the tangent vectors.
We then specify the initial conditions and address initial discontinuities like before:
\begin{lstlisting}[firstline=25,lastline=26]
I.constant_past( [1.0,0.0], time=0.0 )
I.step_on_discontinuities()
\end{lstlisting}
We then integrate up to \(t=1000\) to avoid transients and let the tangent vectors align themselves:
\begin{lstlisting}[firstline=27,lastline=28]
for time in range(100,1000,100):
	I.integrate(time)
\end{lstlisting}
Note that we do not use one big (sampling) step here as the orthonormalizations coincide with the sampling step and should not happen too rarely to avoid numerical over- or underflows.

\texttt{I.integrate} does not only return the state after integration, but also the local Lyapunov exponents during that interval as well as the exact length of integration contributing to these.
This length may differ from the length of the sampling step and should ideally be used as a weight for the respective local Lyapunov exponents (the variation of weights is small in this example though).
During the actual integration, we collect the local Lyapunov exponents and weights in lists (\texttt{lyaps} and \texttt{weights}), while we discard the states.
\begin{lstlisting}[firstline=30,lastline=36]
times = range(1000,100000,100)
lyaps = []
weights = []
for time in times:
	state,lyap,weight = I.integrate(time)
	lyaps.append(lyap)
	weights.append(weight)
\end{lstlisting}
Finally, we apply a weighted average to the local Lyapunov exponents to obtain the global ones and print them:
\begin{lstlisting}[firstline=38,lastline=39]
from numpy import average
print(average(lyaps,axis=0,weights=weights))
\end{lstlisting}
We obtain \(\lambda_1 \approx 3\times 10^{-6}\), \(\lambda_2 \approx -5\times 10^{-3}\), and \(\lambda_3 \approx -5\times 10^{-2}\).
To decide which of these results is actually significantly different from zero, we can employ Student's one-sample \(t\)\nobreakdash-test (ignoring the weights):
\pagebreak
\begin{lstlisting}[firstline=40,lastline=41]
from scipy.stats import ttest_1samp
print(ttest_1samp(lyaps,popmean=0,axis=0))
\end{lstlisting}
We cannot reject the null hypothesis that \(\lambda_1\) is zero (\(p=0.9\)), but for \(\lambda_2\)~and~\(\lambda_3\), we can clearly reject it (\(p=0\) numerically).
This conforms with the expectation of a periodic dynamics for this set of parameters~\cite{somolinos1978} as well as with optical inspection (Fig.~\ref{fig:timeseries}, central panel).

\subsection{Largest transversal Lyapunov exponent}
JiTCODE and JiTCDDE also provide tools that ease the calculation of the maximal Lyapunov exponent transversal to a synchronization manifold~\cite{heagy1994}.
In addition to what is done for Lyapunov exponents (see Sec.~\ref{lyapunov}), the following steps are automatized:
\begin{itemize}
	\item Synchronized dynamical variables are treated as identical and only one representative of a group of synchronized variables is integrated.
		This avoids redundant calculations and that the dynamics escapes from the synchronization manifold.
	\item The tangent vectors and the corresponding differential equations are transformed such that they only cover directions orthogonal to the synchronization manifold (see App.~\ref{transformation} for detail).
		This obviates the need for removing projections to the synchronization manifold from the tangent vector and thus avoids any ensuing inaccuracies or numerical efforts, which may be considerable for DDEs.
\end{itemize}

\subsubsection*{Example: Transversal Lyapunov exponent of two delay-coupled FitzHugh--Nagumo oscillators}

We consider a system of two FitzHugh--Nagumo oscillators that are diffusively delay-coupled twice in each component~\cite{Saha2017}:
\begin{equation}
	\begin{alignedat}{3}\label{eq:FHN}
		\dot{y}_0 &=  y_0(t) ( y_0(t) -1)(a- y_0(t) )- y_1(t)  &~+~ C_{0,2},\\
		\dot{y}_1 &= b y_0(t)  - c y_1(t)            &~+~ C_{1,3},\\
		\dot{y}_2 &=  y_2(t) ( y_2(t) -1)(a- y_2(t) )- y_3(t)  &~+~ C_{2,0},\\
		\dot{y}_3 &= b y_2(t)  - c y_3(t)            &~+~ C_{3,1},\\
	\end{alignedat}
\end{equation}
where the coupling terms are defined as:
\[
	C_{i,j} \defi M_1 \kl{ y_j(t-\tau_1)-y_i(t) }
				+ M_2 \kl{ y_j(t-\tau_2)-y_i(t) },
\]
and the control parameters are 
\(\tau_1 = 80.0\),
\(\tau_2 = 70.0\),
\(M_1 = 0.005\),
\(M_2 = 0.0053\),
\(a = -0.025\),
\(b =  0.00652\), and
\(c =  0.02\).
For these parameters, the dynamics mostly resides very close to the synchronization manifold defined by \(y_0=y_2\) and \(y_1=y_3\), but occasionally exhibits large excursions from it (see Fig.~9 in Ref.~\onlinecite{Saha2017}).
For a better understanding of this dynamics, it is therefore of interest to quantify the transversal stability of the synchronisation manifold.
As the synchronisation manifold is invariant, its stability is reflected by~\(\lambda_\bot\), the largest Lyapunov exponent transversal to this synchronization manifold.
While we here only perform this analysis for an exemplary choice of parameters, it is straightforward to extend this to perform a parameter scan.

As usual, we start with defining the constants:
\begin{lstlisting}[firstline=1,lastline=7]
tau1 = 80.0
tau2 = 70.0
M1 = 0.005
M2 = 0.0053
a = -0.025
b =  0.00652
c =  0.02
\end{lstlisting}
We then define the coupling terms \(C_{ij}\) and the right-hand side of Eq.~\ref{eq:FHN}:
\begin{lstlisting}[firstline=9,lastline=21]
from jitcdde import y, t
def C(i,j):
	return (
			  M1 * ( y(j,t-tau1)-y(i) )
			+ M2 * ( y(j,t-tau2)-y(i) )
		)

twoFN_f = [
		y(0)*(y(0)-1)*(a-y(0))-y(1) + C(0,2),
		b*y(0) - c*y(1)             + C(1,3),
		y(2)*(y(2)-1)*(a-y(2))-y(3) + C(2,0),
		b*y(2) - c*y(3)             + C(3,1),
	]
\end{lstlisting}
We create a list \texttt{groups} that contains tuples specifying which groups of variables are synchronized and use it to intitialize the integrator:
\begin{lstlisting}[firstline=23,lastline=25]
groups = [ (0,2), (1,3) ]
from jitcdde import jitcdde_transversal_lyap
I = jitcdde_transversal_lyap(twoFN_f,groups)
\end{lstlisting}
The remaining procedure is very similar to our previous example for regular Lyapunov exponents (see Sec.~\ref{sunflower_lyap}).
Note that we here control the maximum step used within \texttt{step\char`_on\char`_discontinuities} to avoid numerical issues due to an overly large step size.
Also note that the variability of the weights is again small, allowing us to use an unweighted \(t\)\nobreakdash-test.
\begin{lstlisting}[firstline=27,lastline=43]
I.constant_past([1.0,0.0],time=0.0)
I.step_on_discontinuities(max_step=1)
for time in range(100,1000,100):
	I.integrate(time)

times = range(1000,300000,100)
lyaps = []
weights = []
for time in times:
	state,lyap,weight = I.integrate(time)
	lyaps.append(lyap)
	weights.append(weight)

from numpy import average
print(average(lyaps,weights=weights))
from scipy.stats import ttest_1samp
print(ttest_1samp(lyaps,popmean=0))
\end{lstlisting}
We obtain \(\lambda_\bot \approx 0.0011\), which is non-zero with \(p=10^{-6}\).
Such a positive transversal Lyapunov exponent corroborates the observation that the synchronized state is unstable~\cite{Saha2017}.

\subsection{Further performance-oriented features}

In many typical applications it is desirable to evolve the same or a similar differential equation in parallel, e.g., when investigating the impact of initial conditions or control parameters.
In this case, it is expedient to avoid repeating the generation and compilation of the C~code, as it may take a considerable time.
Therefore, the presented modules offer the functionality to store and load a file containing the compiled code.
Moreover, it is possible to leave control parameters undefined at compile time and only specify them at integration time.

Another possible target of optimization are considerable redundancies amongst the right-hand side of the differential equation.
A typical example is a mean-field coupling, in which case the same expression for the mean field appears in the expressions for all components coupled to it.
Such redundancies can be automatically handled by the compiler's or SymPy's common-subexpression evaluation, but those may take time or may not be able to handle large differential equations.
Therefore, it is also possible for the user to manually define repeating expressions (called \textit{helpers)} and use them in the differential equations.

Finally, the presented modules can optionally employ multiprocessing (using OpenMP~\cite{dagum1998}) to calculate the derivative and for some other suitable and time-consuming operations like calculating the scalar product between separation functions (see App.~\ref{DDE_lyap}).
Due to the inevitable overhead of multiprocessing, this only is worthwhile for larger differential equations.
Moreover, if the number of realizations of some setup that need to be computed is equal to or higher than the number of available cores, it may be more efficient to run these realizations in parallel instead of running parallelized realizations in sequence.

\begin{figure*}
	\includegraphics{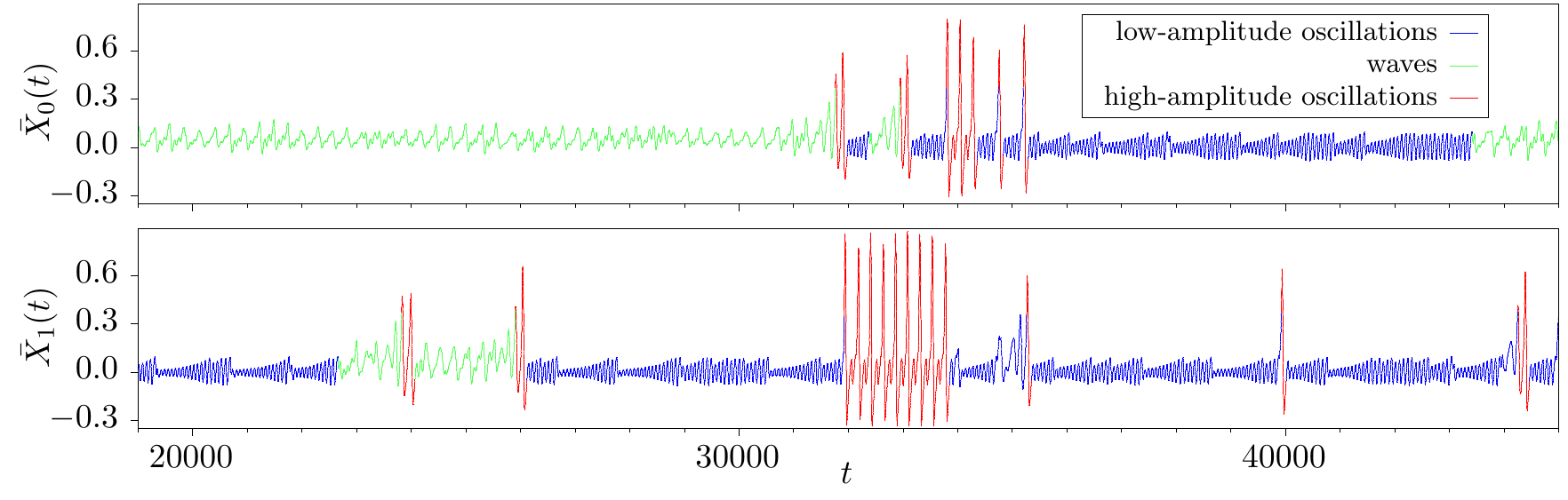}
	\caption{
		Exemplary excerpt of the time series generated by the example program in Sec.~\ref{pattern_hopping}, which integrates two coupled small-world networks of FitzHugh--Nagumo oscillators.
		\(\bar{X}_q \defi \sum_{i=0}^{N-1} X_{iq}\) denotes the mean of the variable~\(X\) of the respective subnetwork.
		Line colors mark dynamical patterns.
	}
	\label{fig:hopping}
\end{figure*}

\subsubsection*{Example: Two coupled small-world networks of FitzHugh--Nagumo oscillators} \label{pattern_hopping}

Following recent studies of networks of networks~\cite{Um2011, Dagostino2014, Kenett2015, Sonnenschein2015}, we consider a system of two identical two-dimensional small-world networks of FitzHugh--Nagumo oscillators.
The dynamics of a single such network has been described and investigated in Ref.~\onlinecite{Ansmann2016}, which found that the dynamics performs self-induced switchings between three different space--time patterns, namely low-amplitude oscillations, traveling waves, and short-lived high-amplitude oscillations (extreme events).
These switchings were found to be part of a long transient which either settles on an attractor corresponding to high-amplitude oscillations or permanent traveling waves.

We couple two such networks to each other so that each oscillator in one subnetwork is weakly coupled to all oscillators in the other network, i.e., the two subnetworks are coupled to each other via their mean field.
The oscillator~\(i\) of subnetwork~\(q\) is described by the following differential equations:
	\begin{equation} \label{eq:fhn_sw}
	\begin{aligned}
		\dot{X}_{iq} & = &&
		X_{iq} (a-X_{iq}) (X_{iq}-1) - Y_{iq}\\
		&& +~ & \hphantom{\sum\limits_{r=0}^{1} B_{qr}}\frac{k_\text{W}}{M} \sum\limits_{j=0}^{N-1} A_{ij} \kl{X_{jq} - X_{iq}}\\
		&& +~ & \sum\limits_{r=0}^{1} B_{qr} \frac{k_\text{B}}{N}  \sum\limits_{j=0}^{N-1}\kl{X_{jr} - X_{iq}}
		, \\
		\dot{Y}_{iq} & = && b_{iq} X_{iq} - c Y_{iq}.
	\end{aligned}
	\end{equation}
Note that we capitalized some variables in comparison to Ref.~\onlinecite{Ansmann2016} for disambiguation.
The control parameters are \(a = -0.0276\), \(b_i \sim \mathcal{U}\kl{\kle{0.006,0.014}}\), and \(c = 0.02\).
Each subnetwork has \(N=L^2\) nodes with \(L=100\), and the coupling \textbf{w}ithin it is governed by the coupling strength \(\frac{k_\text{W}}{M}\) with \(k_\text{W} = 0.128\) and \(M=60\) as well as by the adjacency matrix~\(A \in \klg{0,1}^{n \times n}\), which describes a small-world network with a rewiring probability~\(p=0.18\) based on an \(L \times L\)~lattice with cyclic boundary conditions, where each node is coupled to its \(M\)~nearest neighbours (see Ref.~\onlinecite{Ansmann2016} for details).
The coupling \textbf{b}etween each of the two subnetworks is governed by the coupling strength \(\frac{k_\text{B}}{N}\) as well as by the adjacency matrix
\[B=\begin{pmatrix}0 & 1\\1 & 0\end{pmatrix},\]
which describes a complete coupling.

We start by defining the constants and the values of the inhomogeneous control parameter~\(b\):
\begin{lstlisting}[firstline=1,lastline=9]
L = 100
N = L*L
M = 60
p = 0.18
a = -0.0276
from numpy.random import uniform
b = uniform(0.006,0.014,N)
c =  0.02
k_W = 0.128
\end{lstlisting}
As we want to be able to vary the coupling strength between networks without repeating the compilation, we define the control parameter \(k_\text{B}\) as SymEngine symbol instead of a number:
\begin{lstlisting}[firstline=11,lastline=12]
from symengine import Symbol
k_B = Symbol("k_B")
\end{lstlisting}
We continue with defining the adjacency matrices of the networks, where we employ an external function for generating the small-network (which can be found in the ancillary files):
\begin{lstlisting}[firstline=14,lastline=19]
from small_world_2d import small_world_2d
A = small_world_2d(L,M,p)
B = [
		[ 0, 1 ],
		[ 1, 0 ],
	]
\end{lstlisting}
Before we start defining the actual differential equations, we have to decide which component of \(y\) from Eq.~\ref{eq:ode} represents which component of which oscillator.
We here choose:
\begin{equation} \label{eq:mapping}
\begin{aligned}
	y_0    &= X_{0,0}, &&\ldots, &y_{N-1}  &= X_{N-1,0},\\
	y_N    &= Y_{0,0}, &&\ldots, &y_{2N-1} &= Y_{N-1,0},\\
	y_{2N} &= X_{0,1}, &&\ldots, &y_{3N-1} &= X_{N-1,1},\\
	y_{3N} &= Y_{0,1}, &&\ldots, &y_{4N-1} &= Y_{N-1,1},
\end{aligned}
\end{equation}
and create utility functions that implement this mapping:
\begin{lstlisting}[firstline=21,lastline=23]
from jitcode import y
X = lambda i,q: y(  q*2*N+i)
Y = lambda i,q: y(N+q*2*N+i)
\end{lstlisting}
If we implemented Eq.~\ref{eq:fhn_sw} na\"ively, we would probably spend a lot of runtime on computing the between-network coupling sum:
\[
	      \sum\limits_{j=0}^{N-1}\kl{X_{jr} - X_{iq}}
	=     \sum\limits_{j=0}^{N-1}X_{jr}   - N X_{iq}
	\ifed               S_r               - N X_{iq}.
\]
However, most of this can be avoided by exploiting that \(S_r\) only needs to be computed once per subnetwork.
To this end, we define \emph{helpers} for \(S_0\) and~\(S_1\):
\begin{lstlisting}[firstline=25,lastline=29]
S = [Symbol("S_0"),Symbol("S_1")]
helpers = [
		(S[q],sum(X(i,q) for i in range(N)))
		for q in [0,1]
	]
\end{lstlisting}
We can now use the symbols \texttt{S[0]} and \texttt{S[1]} when defining the right-hand side of the differential equation.
Before the latter is evaluated at runtime, the helpers will be computed once; the result will then be used for the evaluation.

With this, we have everything we need to write a generator function implementing Eq.~\ref{eq:fhn_sw}:
\begin{lstlisting}[firstline=31,lastline=55]
def two_FHN_SWs_f():
	for q in [0,1]:
		for i in range(N):
			dXdt = (
					  X(i,q)
					* (a-X(i,q))
					* (X(i,q)-1.0)
					- Y(i,q)
				)
			coupling_sum_W = sum(
					X(j,q)-X(i,q)
					for j in range(N)
					if A[i,j]
				)
			dXdt += k_W/M * coupling_sum_W
			coupling_sum_B = sum(
					S[r]-N*X(i,q)
					for r in [0,1]
					if B[q][r]
				)
			dXdt += k_B/N * coupling_sum_B
			yield dXdt
		
		for i in range(N):
			yield b[i]*X(i,q) - c*Y(i,q)
\end{lstlisting}
Note that it is crucial that nesting of the loops and thus the order in which the \texttt{yield} statements are called reflects the one we chose in Eq.~\ref{eq:mapping}.
We can now pass this function to \texttt{jitcode} along with our previously defined helpers, the symbol for the free control parameter \(k_\text{B}\) and the system size:
\begin{lstlisting}[firstline=57,lastline=63]
from jitcode import jitcode
I = jitcode(
		two_FHN_SWs_f,
		helpers = helpers,
		n = 4*N,
		control_pars = [k_B],
	)
\end{lstlisting}
We want to use multiprocessing for this huge differential equation, which we enable and control by calling dedicated functions for the processing steps (which were performed automatically in the previous examples):
\begin{lstlisting}[firstline=64,lastline=65]
I.generate_f_C(chunk_size=4*L)
I.compile_C(omp=True)
\end{lstlisting}
The parameter \texttt{chunk\char`_size} controls how many components of the right-hand side form a chunk that is computed together and thus can be optimized by the compiler.
We here choose the number of nodes in four lines of the lattice~(\(4L\)), which is a trade-off between the multiprocessing overhead and boost from compiler optimisation on the one side and the compilation time on the other side.
Finally, instead of using the compiled function directly, we save it to a file, so we can use it in a different script:
\begin{lstlisting}[firstline=66,lastline=66]
I.save_compiled()
\end{lstlisting}

The following is an example for a script that loads this file (\texttt{"jitced.so"}), sets the parameter \(k_\text{B}\) to \(4.3 \times 10^{-4}\), integrates the dynamics, and prints the mean value of~\(X\) for each subnetwork:

\begin{lstlisting}
N = 10000
k_B = 4.3e-4

from jitcode import jitcode
I = jitcode(
		n = 4*N,
		module_location = "jitced.so"
	)
I.set_parameters(k_B)
I.set_integrator("dopri5")

from numpy.random import random
initial_state = random(4*N)
I.set_initial_value(initial_state,0.0)

from numpy import mean
times = range(100000)
for time in times:
	state = I.integrate(time)
	print(
			mean(state[  0:  N]),
			mean(state[2*N:3*N]),
		)

\end{lstlisting}
The output is plotted in Fig.~\ref{fig:hopping} (and a video showing the dynamics of each oscillator is a supplement for the journal's version of the paper).
We observe the typical pattern switching as described in Ref.~\onlinecite{Ansmann2016} and --~despite the small coupling strength between networks~-- interactions between the patterns of different subnetworks:
While groups of high-amplitude oscillations almost only occurred as an attractor in a single network, they now also frequently occur as a pattern with finite duration, and they seem to be able to propagate from one subnetwork to the other (at \(t \approx 31900\) and \(t \approx 33800\)).
It is upon future research to investigate the role of the coupling and the mechanism underlying these phenomena.

\section{Runtime analysis}\label{benchmark}

For a large system of differential equations, the total runtime of the presented modules can be divided into two main components: the preparation time, which primarily contains the generation and compilation of code, and the integration time.
The first time primarily scales with \Order[m]{m}, where \(m\)~is the number of operations needed to compute the derivative~\(f\) (and \(g\) for SDEs).
As \(m = \Order{n}\) or worse, the integration time also scales with \Order[m]{m}; moreover it scales with \Order[T]{T}, where \(T\)~is the total integration time.

To experimentally verify this, to investigate the remaining contributions to the runtime, and for comparison with other software, we employ the Kuramoto network presented in Sec.~\ref{kuramotos} as a benchmark.
As a consequence, \(m\)~can be well approximated by the number of edges (\(0.2 \times n (n-1)\)).
To closer mimic the conditions of a typical long-time use case, such as an investigation of the temporal evolution of synchrony, we increase the sampling interval to 10 time units and the total integration time~\(T\) to 20.000 time units.
Furthermore, we vary the size~\(n\) of the network to investigate its influence on the runtime.

We choose an ODE (and thus JiTCODE) for benchmarking due to the selection of available softwares for ODEs, which particularly allows for a comparison with other softwares using the same integration method, namely DoP853.
These are:
\begin{description}
	\item[PyDSTool\cite{clewley2007}]
		Like the presented modules, this Python module offers just-in-time compilation of the derivative.
		However, by contrast, the differential equations are specified as strings and translated to code using SWIG~\cite{beazley2003}.
		Due to their similarity in structural aspects affecting efficiency, we regard PyDSTool as a yardstick for JiTCODE.
		\pagebreak
	\item[Conedy\cite{Rothkegel2012}]
		This C++-based software with a Python interface is specialized to implementing dynamics on networks.
		By contrast to the presented modules, the network structure and inhomogeneous control parameters are looked up at execution time when evaluating the derivative; individual node dynamics are pre-compiled though (this time is not taken into account in our benchmark as it has to be spent only once per type of node).
		As Conedy takes several measures to ensure efficiency, we consider it an epitome of what can be achieved without metaprogramming.
		Still, we expect it to have longer integration times than JiTCODE due to reasons we elaborated in Sec.~\ref{rationale}.
		Note that for our benchmark we make use of \emph{static edges} (see Sec.~VIII.A of Ref.~\onlinecite{Rothkegel2012}).
\end{description}

For each investigated~\(n\), we consider 100~different realizations of the above setup (i.e., of the network and the eigenfrequencies), which we refer to as \emph{scenarios.}
We employed the following measures to avoid biases:
\begin{itemize}
	\item All scenarios were run sequentially on the same machine.
		Background processes were shut down as far as possible.
		Performance was measured in \emph{CPU time}~\cite{*[{Sec.~3.1.2 in }][] mcgeoch2012}.
	\item Each software was subjected to the same selection of scenarios to avoid bias due to the difficulty of the scenarios.
	\item The softwares were subjected to each scenario in direct succession to avoid a bias from long-time fluctuations of the machine's performance.
		The order in which this happened was randomized for each scenario to avoid biases due to caching and similar.
	\item We use the median for summarizing the data to avoid a bias due to temporary outliers of the machine's performance.
\end{itemize}
The scripts used for the benchmark are ancillary files.

\begin{figure}
	\includegraphics{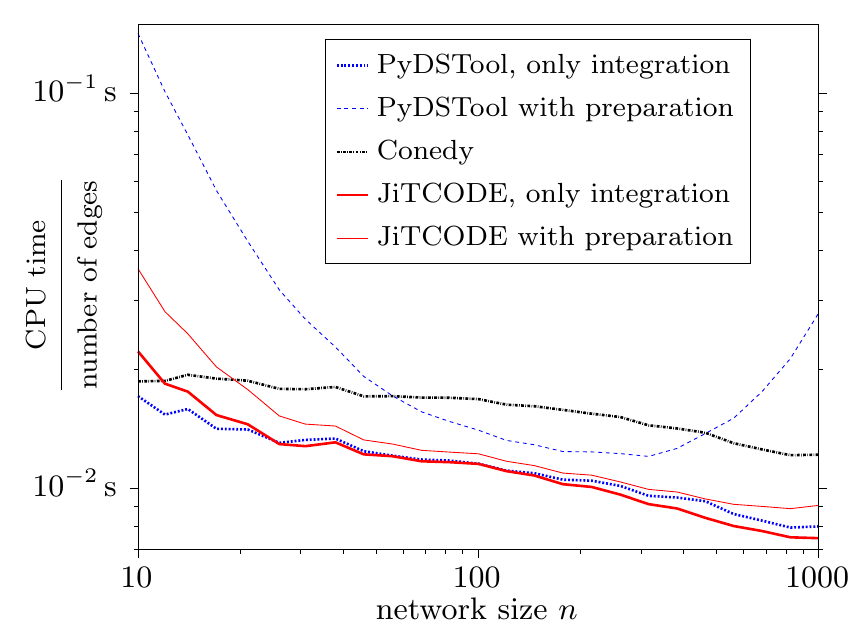}
	\caption{Median runtimes from 100~realizations for different softwares for integrating Eq.~\ref{eq:kuramoto} for \(T=20.000\) time units normalized on the expected number of edges in the network (\(0.2 \times n (n-1)\)).
	Thick lines mark the pure integration times; thin lines mark the entire time needed including preparation steps such as code generation and compilation (for Conedy, the time consumed by such preparation steps is negligible).
	Experiments were performed on a machine with two 2.4\,GHz Intel Xeon E5620 processors and with 24\,GB RAM, using Ubuntu 14.04.5 LTS and the included Clang as a compiler.
	}
	\label{fig:performance}
\end{figure}

In Fig.~\ref{fig:performance}, we show in dependence of the network size~\(n\), the median runtimes normalized by the number of edges (which scale with \Order{n^2}).
That these normalized runtimes fluctuate at most by one order of magnitude confirms our expectation that the unnormalized runtimes primarily depend on the number of operations~\(m\) and thus on the number of edges.
For all softwares, we observe that the normalized integration time slightly decreases with~\(n\), which we attribute to vector arithmetics used by the integrator, which scales with \Order{n} and whose relative impact consequentially decreases with~\(n\).
There is little difference in integration speed between JiTCODE and PyDSTool, while they are roughly 1.6 times as fast as Conedy for large~\(n\), which we attribute to the overhead from iterating over the adjacency list at execution time.
Note that this difference may be larger for other problems such as weighted networks where Conedy cannot benefit from using completely static edges (see Sec.~VIII.A of Ref.~\onlinecite{Rothkegel2012}).

When the preparation time is taken into account, JiTCODE's speed is only comparable to that of PyDSTool for medium~\(n\).
A possible explanation for the big difference for large~\(n\) is that in many settings, compilers are not good at handling large pieces of unstructured code, and JiTCODE introduces an artificial structure to avoid this by default, while we are not aware of such a feature for PyDSTool.
Even, with the preparation time taken into account, JiTCODE is still faster than Conedy for large~\(n\).
Note, however, that this outcome depends on the total integration time~\(T\): for a sufficiently small integration time, Conedy will outperform JiTCODE as its preparation time is negligible.

While we showed and discussed the results for the median CPU time in the two above paragraphs, we obtained qualitatively similar results for the mean, minimum, and maximum as well as for \emph{wall-clock time}~\cite{*[{Sec.~3.1.2 in }][] mcgeoch2012}.

\section{Conclusion}
JiTCODE, JiTCDDE, and JiTCSDE allow to integrate differential equations of different types, combining the ease of use of a higher-level programming language and the efficiency of a lower-level language.
While we started with simple examples for didactic reasons, we also demonstrated that it also allows to solve complex and timely problems, namely integrating networks of networks and determining transversal Lyapunov exponents of systems with multiple delays.
We theoretically argued for the numerical efficiency of the presented approach and exemplarily showed that JiTCODE outperforms comparable modules when integrating a complex network.
We showed the implementation of a network with 20.000 nodes and 1.200.000 edges but have no reason to assume that this represents an upper limit of some sort.

A distinctive property of the presented modules is that the differential equations are specified symbolically.
This allows for an automated flexible processing within the presented modules, which we particularly exploited to implement an automatized computation of regular and transversal Lyapunov exponents, but also use for several smaller convenient features.
It is conceivable to extend this approach to other numerical methods of analysis, such as continuation~\cite{krauskopf2007}.
All the employed integration algorithms use an adaptive integration step size, thus sparing the user the tedious and error-prone process of choosing an appropriate step size themselves as well as providing for an enhanced efficiency for dynamics where the required step size fluctuates strongly over time.
Despite all of these automatisms, users can tweak many parts of the process such as symbolic optimizations as well as compilation and integration parameters.
Also, while these automatisms are convenient and prevent trivial errors, they do not spare the user from the responsibility to understand \emph{what} is automatized, to avoid common pitfalls, and to take care when interpreting the results.

Since the first usable versions, we have seen many successful uses of the presented modules from colleagues, collaborators, and others.
These did not only confirm our general approach of design but also inspired or stimulated new or improved features.
We hope to see more such developments in the future.

\begin{acknowledgments}
	\hyphenation{Freund Leh-nertz Bro-se Ry-din Za-ba-wa Pi-kov-sky Ujj-wal Coel-ho Ver-nie As-lim}
	I am grateful to 
	M.~Anvari,
	D.~Biswas,
	H.~Dickten,
	U.~Feudel,
	J.~Freund,
	C.~Geier,
	J.~Heysel,
	K.~Lehnertz,
	A.~Pikovsky,
	C.~Rackauckas,
	M.R.~Rahimi Tabar,
	L.~Rydin Gorj\~ao, and
	A.~Saha
for constructive discussions, to
	E.~Aslim,
	D.~Biswas,
	A.~Blendl,
	J.~Brose,
	F.~Code\c{c}o Coelho,
	H.~Dickten,
	C.~Geier,
	R.~Gommers,
	M.~Roskosch,
	T.~Rings,
	E.~Roque,
	L.~Rydin Gorj\~ao,
	A.~Saha,
	S.~Ujjwal,
	J.~Vernie,
	Y.~Wang,
	H.~Yao, and
	L.~Zabawa,
for testing and reviewing the software and documentation, as well as to
	N.~Ansmann,
	C.~Geier,
	K.~Lehnertz,
	A.~Saha, and
	K.~Simon
for constructive comments on earlier versions of the manuscript.
This work was supported by the Volkswagen Foundation (Grant No.~88463)
\end{acknowledgments}

\pagebreak
\appendix
\section{Efficiently storing and accessing past states in an adaptive DDE solver}\label{past_access}

The Shampine--Thompson method stores and accesses past states as follows:
After each integration step, the current time, state, and derivative (which is a by-product of the underlying Bogacki--Shampine integrator) are stored as what we refer to as an \textit{anchor.}
If a past state is required for evaluating the derivative~\(f\), the neighboring anchors are located and cubically Hermite-interpolated to obtain the past state.

If the integration steps were equidistant, the anchors could be stored in a circular buffer (whose length is the maximum delay divided by the step size) and the location of the needed anchors could be straightforwardly computed from the delay and the buffer's state.
However, as the step size is adaptive, storage and access of anchors must be more sophisticated to avoid an unnecessary loss of performance.
We here store the anchors in a doubly linked list, as this structure is not restricted in size and allows adding and removing anchors individually without frequently rebuilding the entire structure (which would need to be done in case of arrays, for example).
To quickly find anchors, we make use of the fact that for a given delay term \(y(t-\tau)\), the required past times \(t-\tau\) only change slowly during the integration.
Due to this, we can efficiently locate the needed anchor by traversing the list and storing the search position \textit{(cursor)} between different searches.
This way, the number of anchors we need to traverse for each individual evaluation of a delay term is one or less in most cases.

As finding the right pair of anchors and obtaining the actual state are separated on the symbolic level, searching the same anchors twice in case of a duplicate delay can be avoided by the user or automatically through a common-subexpression elimination on the symbolic level.

\section{Scalar products of separation functions with an adaptive DDE solver}\label{DDE_lyap}

\newcommand{\taumax}{\ensuremath{\max_j \tau_j}}
To calculate the Lyapunov exponents of DDEs, we cannot only consider tangent vectors, but also have to take into account their pasts~\cite{farmer1982} --~called \textit{separation functions}~-- up to the maximum delay~\taumax, as they correspond to the full state of the system.

Farmer~\cite{farmer1982} considers the case of a fixed integration step size~\(\Delta t\) that divides the only delay~\(\tau\).
He represents a separation function as a vector consisting of its values at the integration steps and uses the standard scalar product between two separation functions \(v\) and~\(w\) for purposes of obtaining norms and (ortho)normalizing:
\[
	\kls{v,w}
	= \sum_{i=0}^\frac{\tau}{\Delta t}
		v\kl{t-i\Delta t} \,
		w\kl{t-i\Delta t}
.\]
This does not translate well to an adaptive step size as it would not take into account the changing number and distance of sampling points.

Instead, we employ a function scalar product between the piecewise cubic Hermite interpolants (denoted by \(\mathcal{H}\)) of the anchors, which are already stored and used for obtaining past states (see App.~\ref{past_access}):
\[
	\kls{v,w}
	= \int\limits_{t-\taumax}^{t}
		\mathcal{H}_v(\tilde{t}) \cdot
		\mathcal{H}_w(\tilde{t})
		\diff \tilde{t}
.\]

As the interpolants are polynomials, we can evaluate this integral analytically and represent the scalar product as a vector--matrix--vector multiplication in the vector space defined by the stored anchors.
The respective matrix is sparse and was largely calculated beforehand, allowing for an efficient evaluation of this scalar product.

Note that in the limit of an infinitely fine step size, both scalar products are equivalent.

\section{A brief introduction to generator functions} \label{generators}
In addition to lists (and other iterables) of expressions, the presented modules also accept as input generator functions that yield expressions.

A simple example for a generator function is:
\begin{lstlisting}[firstline=14,lastline=17]
def roessler_f():
	yield -y(1) - y(2)
	yield  y(0) + a*y(1)
	yield b + y(2)*(y(0)-c)
\end{lstlisting}
This function returns a generator object when being called.
When this object is iterated over, the function will be executed until the first \texttt{yield} statement is reached and its argument (here: \texttt{-y(1) - y(2)}) is returned as the first element of the iteration.
When the second element is requested, the function is further executed until the next \texttt{yield} statement is reached and its argument (here: \texttt{y(0) + a*y(1)}) is returned as the second element.
This continues until the function is completely executed.
Thus iterating over the output of the above generator function is equivalent to iterating over the following list:
\begin{lstlisting}[firstline=7,lastline=11]
roessler_f = [
		-y(1) - y(2),
		 y(0) + a*y(1),
		b + y(2)*(y(0)-c)
	]
\end{lstlisting}
In the first example (Sec.~\ref{roessler}), we passed this list as an argument to \texttt{jitcode}.
We could just as well have passed the above generator function.

Like regular functions, generator functions can contain control structures such as loops --~we use this in the example in Sec.~\ref{kuramotos}.
While regular iterables store all their elements simultaneously, generators only generate the elements on demand.
In particular for huge differential equations, this may considerably reduce the required memory.
However, the length of a generator can only be determined by iterating over it, which is why it is prudent to specify the dimension of the differential equation explicitly when providing them as a generator function.

\section{Transformations for estimating the transversal Lyapunov exponent} \label{transformation}

We first consider the case that we investigate the stability of the synchronization manifold described by \(y_0 = y_1 = \ldots = y_n\), i.e., there is only one group of synchronized dynamical variables comprising the entire dynamical system.
Furthermore, we assume an ODE for the sole sake of notational simplicity.
Finally, let the differential equation describing the evolution of a tangent vector be:
	\[ \dot{z}(t) = h(t,y(t),z(t)),\]
with
\(z: \mathbb{R} \rightarrow \mathbb{R}^n\),
\(h: \mathbb{R} \times \mathbb{R}^n \times \mathbb{R}^n \rightarrow \mathbb{R}^n\), and
\(y\)~being the solution of the main ODE, i.e., Eq.~\ref{eq:ode}.

\(\mathfrak{z}_0(t) \defi z_0(t) + z_1(t) + \ldots + z_n(t)\) describes the evolution of infinitesimal distances along the synchronization manifold --~corresponding to the basis that contains only the vector \(\kl{1,\ldots,1}\).
To easily separate growths of the tangent vector that are orthogonal to these, we need a basis of the orthogonal complement of the span of that basis.
The following form a sparse basis meeting these requirement:
\begin{equation*}
	\begin{alignedat}{3}
		\mathfrak{z}_1(t) &\defi~& z_0(t) &- z_1(t), \\
		\mathfrak{z}_2(t) &\defi~& z_1(t) &- z_2(t), \\
		&&\vdots \\
		\mathfrak{z}_n(t) &\defi~& z_{n-1}(t) &- z_n(t), \\
	\end{alignedat}
\end{equation*}
Note that there is no need for normalization as we are only interested in the growth of these.
We can write this transformation as a matrix--vector multiplication:
\[
	\mathfrak{z}(t) = A \cdot z(t) \defi
	\begin{pmatrix*}[r]
	1 &  1 &     1  & \cdots &  1  \\
	1 & -1 &        &        &     \\
	  &  1 &    -1  &        &     \\
	  &    & \ddots & \ddots &     \\[5pt]
	  &    &        &    1   & -1  \\
	\end{pmatrix*} \cdot z(t).
\]
The inverse transformation is \(z(t) = A^{-1} \cdot \mathfrak{z}(t)\) with:
\[
\arraycolsep=6pt
	A^{-1} = \frac{1}{n}\; \begin{pmatrix*}[r]
	   1   &   n-1  &   n-2  &   n-3  & \cdots &    1   \\
	   1   &    -1  &   n-2  &   n-3  & \cdots &    1   \\
	   1   &    -1  &    -2  &   n-3  & \cdots &    1   \\
	\vdots & \vdots & \vdots & \ddots & \ddots & \vdots \\
	   1   &    -1  &    -2  & \cdots &  2-n   &    1   \\
	   1   &    -1  &    -2  & \cdots &  2-n   &  1-n   \\
\end{pmatrix*}.
\]
Taking everything together, we can write the differential equation for the transformed tangent vectors:
\[\dot{\mathfrak{z}}(t) = A \cdot h \Big( t,\, y(t),\, A^{-1}\cdot \mathfrak{z}(t) \Big)\]

What we gained with this procedure is that we can now easily pin the component along the synchronization manifold~(\(\mathfrak{z}_0\)) to zero and do not need to numerically remove projections during the integration.

In case of multiple groups of synchronized dynamical variables, we have a basis of the synchronization manifold that contains, for each group, the vector whose elements are \(1\) for the components belonging to that group and \(0\)~otherwise.
The transformation matrix and its inverse are then structured into blocks of the above forms.

\end{document}